\documentstyle[preprint,prc,aps]{revtex}
\begin{document} 
\draft
\title{Enhanced T-odd P-odd Electromagnetic Moments
in Reflection Asymmetric Nuclei}
\author{V. Spevak$^{(1)}$, N. Auerbach,$^{(1)}$ and V.V. Flambaum$^{(2)}$}
\address{$^{(1)}$School of Physics and Astronomy, Tel Aviv University,
Tel Aviv 69978, Israel}
\address{$^{(2)}$School of Physics, University of New South Wales,
Sydney, 2052, NSW, Australia}
\date{December 15, 1996}
\maketitle
\begin{abstract}
Collective P- and T- odd moments produced by parity and time
invariance violating forces in reflection asymmetric nuclei are
considered.  The enhanced collective Schiff, electric dipole and
octupole moments appear due to the mixing of rotational levels of
opposite parity.  These moments can exceed single-particle moments
by more than two orders of magnitude.  The enhancement is due to the
collective nature of the intrinsic moments and the small energy
separation between members of parity doublets.  In turn these
nuclear moments induce enhanced T- and P- odd effects in atoms and
molecules.  First a simple estimate is given and then a detailed
theoretical treatment of the collective T-, P- odd electric moments
in reflection asymmetric, odd-mass nuclei is presented and various
corrections evaluated.  Calculations are performed for octupole
deformed long-lived odd-mass isotopes of Rn, Fr, Ra, Ac and Pa and
the corresponding atoms.  Experiments with such atoms may improve
substantially the limits on time reversal violation.
\end{abstract}
\pacs{21.10.Ky, 21.60.Ev, 24.80.+y, 32.80.Ys}

\narrowtext

\baselineskip=3.3ex

\section{INTRODUCTION}
\label{sec:introduc}

In 1964 Christenson {\em et al.} \cite{Christenson-cp} have
discovered that CP is violated in the decay of neutral K-mesons. As
one expects the CPT theorem to be valid this discovery implied that
time reversal (T) is violated in the observed decays of the
Kaon. This fact has immediately lead to the search of CP or
T-violation in other systems. Since the 60$^{^,}$s many attempts have
been made to observe T-violation in systems different from the Kaons.
Time reversal violation has not been observed so far but upper limits
for T-conservation have been established. The search of time reversal
violation encompasses a large variety of physical systems and
involves many methods. One of the more widely used methods involves
the search for static \mbox{T-,} P- odd electromagnetic moments, it
is moments that would be absent if the Hamiltonian of the system is
even under time reversal and reflection. Such moments include the
electric dipole moment (EDM), the electric octupole, the magnetic
quadrupole etc. Early on, with the discovery of CP violation,
attempts were made to measure the electric dipole moment of the
neutron and at present significant upper limits exist on the
existence of such moment \cite{Smith-neutr-edm,Altarev-neutr-edm}.
The neutron was not the only system in which attempts were made to
find a static electric dipole moment. Experiments with atoms and
molecules were performed in which upper limits for electric dipole
moments of the respective systems were established. In fact the
recent measurements of dipole moments of Hg and Xe atoms
\cite{d-Hg-95}  and TlF molecule
\cite{d-TlF-89} have established upper limits for time reversal
violating nucleon-nucleon and quark-quark interactions
that are of the same order (or maybe even exceed) the
limits obtained in the measurement of the neutron dipole moment.

The existence of a static atomic dipole moment may be due to the
following three reasons: {\em a)}~the possible existence of a dipole
moment of the electron (the best limits on the electron EDM were
obtained in Tl atom EDM measurements in Ref.\ \cite{Tl}, see also Cs
measurements in Ref.\ \cite{Cs}) {\em b)}~time reversal violation in
the electron-nucleon interaction, thus in the lepton-hadron
interactions, {\em c)}~the possible existence of a static T-odd, P-odd
moment of the nucleus arising from the time reversal violating
component of the hadron-hadron interactions.  The recent experiments
with Hg gave the best limit on this interaction.  This possibility
will also be the subject of this work.

In a recent paper \cite{AFS-moments} we put forward a suggestion that
rotating nuclei that have static octupole deformations when viewed in
their intrinsic (body) frame of reference will have enhanced T-odd,
P-odd (for short \mbox{T-,} P- odd) moments if a time reversal and
parity violating interaction is present in the nuclear Hamiltonian.
In the intrinsic frame the nucleus with an octupole deformation has
large octupole, Schiff and dipole moments.  An orientation of these
moments is connected to a nuclear axis $\bbox{n}$ (e.g.\ the dipole
moment is $\bbox{d}= d\bbox{n}$).  In a stationary rotational state
the mean orientation of the axis vanishes 
($\langle \bbox{n}\rangle=0$) since the only possible correlation
\mbox{$\langle\bbox{n}\rangle \propto \bbox{I}$} violates time
reversal invariance and parity (here \bbox{I} is the total angular
momentum of the system). Therefore, the mean values of electric
dipole, octupole and Schiff moments vanishes in laboratory frame if
there is no \mbox{T-,} P- violation. In the nuclei with the octupole
deformation and non-zero intrinsic angular momentum there are doublets
of rotational states of opposite parity with the same angular momentum
\bbox{I} (in molecular physics this phenomenon is called $\Lambda$ -
doubling). A \mbox{T-,} P- odd interaction mixes these rotational
levels. As a result the nuclear axis becomes oriented along the total
angular momentum, 
\mbox{$\langle\bbox{n}\rangle \propto \alpha \bbox{I}$} 
where $\alpha$ is the mixing coefficient.  Due to this
orientation of the nuclear axis by a \mbox{T-,} P- odd interaction the
mean values of the \mbox{T-,} P- odd moments are not zero in the
laboratory frame, e.g.\
$\langle\bbox{d}\rangle=
$\mbox{$d \langle\bbox{n}\rangle \propto \alpha d \bbox{I}$}.

We find two basic enhancement factors in this mechanism: firstly, in
the intrinsic frame the nucleus with an octupole deformation will have
large octupole, Schiff, or dipole electric moments because a large
number of nucleons will contribute to the moments, and secondly due to
the appearance of closely spaced parity doublets in the spectrum of
the nucleus with octupole deformation.  It is not only that the
spacing between the members of the doublets is small but also
(\mbox{T-,} P- odd) interaction will mix well such two states. The
enhanced nuclear Schiff moments that result in such nucleus with a
reflection asymmetric shape will induce \mbox{$\sim 1000$} times enhanced
atomic electric dipoles, and measurements performed with such atoms
may improve upper limits for time reversal violation.

It is the aim of the present work to examine in detail the
consequences of the intrinsic reflection asymmetry on the \mbox{T-,}
P- odd electromagnetic moments in nuclei produced by \mbox{T-,} P- odd
components in the nuclear force and on the induced \mbox{T-,} P- odd
moments in the corresponding atoms.  In this paper we extend the work
in Ref.\ \cite{AFS-moments} attempting to provide an improved and more
detailed theory of the nuclei with asymmetric shapes and of the
resulting \mbox{T-,} P- odd moments if parity time reversal is
violated to some degree in the nuclear force.

Present experimental studies of nuclei in the actinide region (Z
around 88 and N around 134) indicate that these nuclei posses
octupole shapes in the ground state (g.s.)
\cite{Aberg-rev,Jain-Sheline-RMP,Ahmad-Butler,Butler-Nazarewicz-RMP}.
In these nuclei near the
Fermi energy, orbital pairs are coupled strongly by the
octupole-octupole part of the effective nuclear interaction. The
existence of octupole deformations in the actinide nuclei is
manifested in the existence of parity doublet states and parity
doublet bands. The E1 and E3 transitions between these states are
relatively strong, of the order of a Weisskopf unit.  
These experimental findings are supported by theoretical studies.
Some isotopes of Rn, Fr, Ra, Ac, Th, and Pa in the 218$<$A$<$230
region are predicted theoretically to be reflection asymmetric in the
g.s.  The phenomena of octupole instability are observed and
described theoretically also around $Z\sim 56$, $N\sim
88$. Several isotopes of Ba, Ce, Nd, and Sm in the A=140--152
region are known to be octupole-soft and develop reflection
asymmetric shapes at higher spins but no experimental data at present
can confirm such shapes in the g.s.\
\cite{Ahmad-Butler,Butler-Nazarewicz-RMP}.

We will present in this work results for relatively long-lived
neutron-odd isotopes of Rn, Ra, and proton-odd isotopes of Fr, Ac,
and Pa for which there is theoretical and in most cases experimental
evidence of reflection-asymmetric intrinsic shapes in the g.s.

We use the results of the Nilsson- Strutinsky mean field calculations
\cite{Cwiok-Nazarewicz} and employ the Leander {\em et al.}
\cite{Leander-Sheline} particle plus core model to calculate the wave
functions, energy splittings and \mbox{T-,} P- odd interaction matrix
elements between members of the parity doublets. In the calculation of
the intrinsic dipole, Schiff and octupole electric moments we use a
two-liquid drop model.  Strutinsky corrections and corrections due to
pairing are taken into account in this work.

In the past one conjectured that \mbox{T-,} P- odd
moments will be enhanced in the cases when close to the g.s.\ there
is, for whatever reason, a level with the same spin and opposite
parity 
\cite{GFeinberg,Coveney-Sandars,Haxton-Henley,Sushkov-Flamb-Khripl}.
The connection between enhanced E1 transitions between these levels
and the nuclear electric dipole moment was made in the work of Haxton
and Henley \cite{Haxton-Henley}.  
At that time, nuclei which possess an intrinsic reflection
asymmetry and quadrupole deformed nuclei without such an asymmetry
but in which an accidentally close ``parity doublet'' exists,
were treated on an equal footing \cite{Haxton-Henley}.  
The fact that in some nuclei systematically enhanced E1 matrix
elements are the direct consequence of intrinsic reflection asymmetry
was  realized and investigated extensively by Leander {\em et al.}
several years later \cite{Leander-Nazarewicz-Bertsch}.

In the work \cite{Sushkov-Flamb-Khripl} the Schiff moment induced by
nuclear \mbox{T-,} P- odd forces was introduced in the presently used
form and calculations for Xe, Hg, Tl and other interesting cases were
done.  The calculations of the atomic electric dipole moments induced
by the nuclear \mbox{T-,} P- odd Schiff moments were also presented
(we should note that similar considerations for \mbox{T-,} P- odd
effects in molecules and atoms induced by the proton electric dipole
moment were applied in \cite{Sandars} and \cite{Khriplovich1976}).

In our earlier paper \cite{AFS-moments} for the first time the
connection was made between the collective \mbox{T-,} P- odd electric
moments in the {\em intrinsic frame} of reference in
reflection-asymmetric nuclei and these moments in the {\em laboratory}
frame.  It is in this case, due to the collective nature of the
intrinsic moments and the nearly identical intrinsic structure of the
parity doublets one that hopes the electric \mbox{T-,} P- odd moments
will be maximal.

After the present introduction in Sec.\ \ref{sec:edm-schiff-mom} we
define the \mbox{T-,} P- odd moments, including the Schiff moment.
In Sec.\ \ref{sec:schematic} we present a simple expression for
a \mbox{T-,} P- odd moment in the case of a deformed rotating nucleus
in the presence of a \mbox{T-,} P- odd interaction. In the same
section we bring a simple schematic estimate of a Schiff moment in an
octupole deformed nucleus. The first part of Sec.\ \ref{sec:models}
deals with the calculation of the dipole, Schiff and octupole
intrinsic moments in a two-fluid liquid drop model.  In the second
part of this section we present the particle+core model and describe
the calculation of the \mbox{T-,} P- odd matrix elements and mixing
amplitudes. In Sec.\ \ref{sec:calsect} the numerical results are
presented for each of the nuclei and at the end of this section
results are given for the atomic dipole moments. In the last section
(Sec.\ \ref{sec:summary-pt}) a summary is presented.

\section{ATOMIC ELECTRIC DIPOLE  AND NUCLEAR T-, P- ODD MOMENTS}
\label{sec:edm-schiff-mom}

We start from the electrostatic potential of a nucleus
screened by the electrons of the atom. 
If the nucleus has a  \mbox{T-,} P- odd dipole moment the
dipole term in the potential vanishes in accordance
with the  Purcell-Ramsey-Schiff theorem \cite{Purcell-Ramsey,Schiff}
(see the derivation in Appendix \ref{app:schiff-theorem}):
\begin{equation}
\varphi(\bbox{R})= \int \frac{e \rho(\bbox{r})}
{\vert \bbox{R}-\bbox{r}\vert}d^{3}\bbox{r}
+\frac{1}{Z}(\bbox{d}\bbox{\nabla})
\int \frac{\rho(\bbox{r})}{\vert \bbox{R}-\bbox{r}\vert}d^{3}\bbox{r} \ .
\label{fb1} 
\end{equation}
Here $\rho(\bbox{r})$ is the nuclear charge density, 
\mbox{$\int \rho(\bbox{r})d^{3}\bbox{r} =Z$}, and
\mbox{$\bbox{d}= \int e \bbox{r}\rho(\bbox{r})d^{3}\bbox{r}$}
is the electric dipole moment (EDM) of the nucleus.
The first term in this
expression is usual electrostatic nuclear potential, and 
the second term is a result of the electron screening effect.
The multipole expansion of $\varphi(\bbox{R})$ contains both 
\mbox{T-,} P- even and \mbox{T-,} P- odd terms. 
We consider here only the latter. 
The dipole part in Eq.\ (\ref{fb1}) is canceled by the second term in
this  equation: 
\begin{equation}
- \int e (\bbox{r}\bbox{\nabla}\frac{1}{R})\rho(\bbox{r})d^{3}\bbox{r}
+\frac{1}{Z}(\bbox{d}\bbox{\nabla})\frac{1}{R}
\int \rho(\bbox{r})d^{3}\bbox{r} = 0  \ .
\label{fb2} 
\end{equation}
The next term is the electric quadrupole which is \mbox{T-,} P- even,
thus the first non zero \mbox{T-,} P- odd term is:
\begin{eqnarray}
\varphi^{(3)}=&& - \frac{1}{6}
\int e \rho(\bbox{r}) r_{\alpha}r_{\beta}r_{\gamma}d^{3}\bbox{r}
\nabla_{\alpha}\nabla_{\beta}\nabla_{\gamma}\frac{1}{R}
\nonumber \\
&&
+\frac{1}{2Z}(\bbox{d}\bbox{\nabla})
\nabla_{\alpha}\nabla_{\beta}\frac{1}{R}
\int \rho(\bbox{r})r_{\alpha}r_{\beta}d^{3}\bbox{r}  \ .
\label{fb3} 
\end{eqnarray}

Here \mbox{$r_{\alpha}r_{\beta}r_{\gamma}$} is a reducible tensor.
After separation of the trace there will be terms which will contain
a vector $\bbox{S}$ 
and a rank 3 tensor $Q_{\alpha\beta\gamma}$ 
(see e.g.\ \cite{Sushkov-Flamb-Khripl}):
\begin{eqnarray}
&& \varphi^{(3)}=\varphi_{\text{Schiff}}^{(3)}+
\varphi_{\text{octupole}}^{(3)} \ ,
\nonumber \\
&& \varphi_{\text{Schiff}}^{(3)}=
-\bbox{S}\bbox{\nabla}\Delta\frac{1}{R}=
4\pi \bbox{S} \bbox{\nabla}\delta(R) \ ,
\nonumber \\
&&\varphi_{\text{octupole}}^{(3)}= 
-\frac{1}{6}Q_{\alpha\beta\gamma}
\nabla_{\alpha}\nabla_{\beta}\nabla_{\gamma}\frac{1}{R}  \ ,
\label{octup-field}
\end{eqnarray}
where
\begin{equation}
\bbox{S} = \frac{1}{10} \biggl (
\int e\rho(\bbox{r}) r^{2}\bbox{r}d^{3}\bbox{r} -
\frac{5}{3}\bbox{d}\frac{1}{Z}\int\rho(\bbox{r})r^{2}
d^{3}\bbox{r} \biggr ) 
\label{fb5} 
\end{equation}
is the Schiff moment (SM) and  
\begin{eqnarray}
&&Q_{\alpha\beta\gamma}
=\!\int e\rho(\bbox{r})
[r_{\alpha}r_{\beta}r_{\gamma}-\frac{1}{5}
(\delta_{\alpha\beta}r_{\gamma} +\delta_{\beta\gamma}r_{\alpha} +     
\delta_{\alpha\gamma}r_{\beta})] d^{3}\bbox{r}
\nonumber \\
&&Q_{zzz} \equiv \frac{2}{5}Q_{3} =
\frac{2}{5}\sqrt{\frac{4\pi}{7}}
\int e\rho(\bbox{r})r^{3}Y_{30} d^{3}\bbox{r}
\label{q-zzz-y30}
\end{eqnarray}
is the electric octupole moment. 
Because the intrinsic dipole moment of the nucleus appears in
second order in the nuclear deformation (see Eq.\ (\ref{dipmom}))
a correction to the octupole field which arises from the non-spherical
part of the density in the screening term in Eqs. (\ref{fb3},\ref{fb5})
can be neglected. Indeed, this correction is at least third order in
the nuclear deformation.

In the absence of T- and P- violating interactions the electric dipole
moment of an atom is equal to zero.  The interaction between atomic
electrons and the \mbox{T-,} P- odd part of the electrostatic nuclear
potential in Eq.\ (\ref{octup-field}) will mix atomic states of the
opposite parity and thus generate an atomic electric dipole moment:
\begin{equation}
D_z = -e \langle \widetilde{\psi} | r_z | \widetilde{\psi} \rangle
= -2 e \sum_{| k_2 \rangle} \frac{\langle k_1 | r_z | k_2 \rangle
\langle k_1 | -e \varphi^{(3)} | k_2 \rangle}{E_{k_1} - E_{k_2}} \ ,
\label{eaedm3a}
\end{equation}
where $\widetilde{\psi}$ denotes the perturbed atomic wave function,
$\mbox{$| k_1 \rangle$} =\mbox{$|k_1,j_1,j_{1z}\rangle$}$ 
is the unperturbed electron ground state
and $\{| k_2 \rangle \}$ is the set of opposite parity states with
which $| k_1 \rangle$ is mixed due the perturbation 
\mbox{$ -e \varphi^{(3)}$}.

The most accurate measurements of atomic and molecular \mbox{T-,} P-
odd electric dipole moments have been done in the atoms Xe and Hg with
zero electron angular momentum, $j_1=0$.  Examining 
Eq.\ (\ref{eaedm3a}) it is easy to demonstrate that in such atoms
nuclear electric octupole (as well as another \mbox{T-,} P- odd
moment, magnetic quadrupole) cannot generate an atomic electric
dipole.
Indeed, according to the triangle rule for the addition of angular
momenta, \mbox{$\langle k_1 | r_z | k_2 \rangle$} can only have 
a nonzero value if 
$\mbox{$|j_1 - j_2|$} \le 1 \le \mbox{$j_1 + j_2$}$. 
Similarly, for
\mbox{$\langle k_1 |\varphi_{\text{octupole}}^{(3)} | k_2 \rangle$}
to be nonzero, we must have
$\mbox{$|j_1 - j_2|$} \le 3 \le \mbox{$j_1 + j_2$}$. This implies
that the following conditions need to be satisfied
for the dipole moment to be nonzero:
\begin{equation}
|j_1 - j_2| \le 1 \text{~ and ~} j_1 + j_2 \ge 3 \ .
\label{eaedm3b}
\end{equation}
The lowest pair of values that satisfies this condition is $j_1 = 3/2$
and $j_2 = 3/2$ for the states with one or odd number of electrons
outside the closed subshells, and $j_1=1$, $j_2=2$ for states with
even number of electrons. In the case of the magnetic quadrupole one
needs $\mbox{$j_1+j_2$} \ge 2$, i.e.\ the lowest pair is $j_1=1/2$,
$j_2=3/2$. Hence, the nuclear electric octupole and magnetic
quadrupole moments cannot contribute
to the atomic electric dipole moment of $j_1=0$ states. 
Therefore, we mostly center our considerations on the Schiff moment.
Note also, that the \mbox{T-,} P- odd part of the electrostatic
nuclear potential in Eq.\ (\ref{octup-field}) is concentrated mainly
inside the nucleus. As a result the induced atomic electric dipole
moment is proportional to the density of external electrons at the
nucleus (more accurately, to the gradient of this density) which
rapidly increases with the nuclear charge $Z$. This is why in general
heavy atoms and nuclei are favored in the studies of \mbox{T-,} P- odd
moments.

\section{SIMPLE ANALYTICAL ESTIMATE OF THE T-, P-ODD MOMENTS}
\label{sec:schematic}

\subsection{T-, P- odd moments and rotational doublets}
\label{subsec:rotat-dublets}

If a deformed nucleus in the {\em intrinsic} (body-fixed) frame of
reference is reflection asymmetric, it can have collective \mbox{T-,}
P- odd moments.  As a consequence of this reflection asymmetry,
rotational doublets appear in the laboratory system. Without
\mbox{T-,} P- violating forces a \mbox{T-,} P- odd moment vanishes
exactly in the laboratory.  A \mbox{T-,} P- odd interaction however
may reveal such intrinsic \mbox{T-,} P- odd moments in the laboratory
frame.  Consider a nearly degenerate rotational parity doublet in the
case of an axially symmetric nucleus.  The wave functions of the
members of the doublet are written as \cite{Bohr-Mottelson}
\begin{equation}
\Psi^{\pm}=\frac{1}{\sqrt{2}}
(\vert IMK \rangle \pm \vert IM-K \rangle) \ .
\label{fb6} 
\end{equation}
Here $\bbox{I}$ is the nuclear spin, \mbox{$M=I_{z}$} and
$K=\bbox{In}$, where $\bbox{n}$ is a unit vector along the nuclear
axis. 

The {\em intrinsic} dipole and Schiff moments are directed along 
$\bbox{n}$: 
\begin{eqnarray}
\bbox{d}_{\text{intr}}=d_{\text{intr}}\bbox{n} \ ,
\nonumber \\
\bbox{S}_{\text{intr}}=S_{\text{intr}}\bbox{n} \ .
\label{simple-schiff-dipole} 
\end{eqnarray}

For these good parity states 
\mbox{$\langle \Psi^{\pm} | \bbox{I} \bbox{n} | \Psi^{\pm}
\rangle$}$= 0$
because $K$ and $-K$ have equal probabilities
and this means that there is no
average orientation of the nuclear axis in the laboratory frame
\mbox{($\langle \Psi^{\pm} | \bbox{n} | \Psi^{\pm} \rangle$}$=0$).
This is a consequence of time invariance and parity
conservation since
the correlation $\bbox{I}\bbox{n}$ is T-, P-odd. As a result
of \mbox{$\langle \Psi^{\pm} | \bbox{n} | \Psi^{\pm} \rangle$}$=0$,
the mean value of the \mbox{T-,} P- odd moments
(whose orientation is determined by the direction of the
nuclear axis)
is zero in the laboratory frame.  

A \mbox{T-,} P- odd interaction $H_{TP}$ will mix the
members of the doublet.
The admixed wave function of the predominantly positive parity
member of the doublet will be
\mbox{$\Psi =\Psi^{+}+\alpha\Psi^{-}$} or
\begin{equation}
\Psi =\frac{1}{\sqrt{2}}
\Bigl ((1+\alpha)\vert IMK \rangle+ (1-\alpha)\vert IM-K \rangle
\Bigr) \ ,
\label{fb6-a} 
\end{equation}
where $\alpha$ is the \mbox{T-,} P- odd admixture
\begin{equation}
\alpha = \frac{\langle \Psi^- | H_{TP} |\Psi^+ \rangle}{E^{+}-E^{-}} \ ,
\label{enad32}
\end{equation}
and $E^{+}-E^{-}$ is the energy splitting between the members of the
parity doublet.
A similar expression is obtained for the negative parity
member of the doublet.

In the \mbox{T-,} P- admixed state
\begin{equation}
\langle \Psi | \bbox{I} \bbox{n} | \Psi \rangle
= \langle \Psi | \hat{K} | \Psi \rangle
= 2 \alpha K \ ,
\label{eceom2d}
\end{equation}
i.e.\ the nuclear axis $\bbox{n}$
is oriented along the nuclear spin $\bbox{I}$:
\begin{equation}
\langle \Psi \vert n_{z} \vert \Psi \rangle =
2\alpha \frac{KM}{(I+1)I} \ .
\label{fb7} 
\end{equation}
Therefore in the laboratory system the electric dipole and Schiff
moments obtain nonzero average values.
For example in the ground state (g.s.) usually \mbox{$M=K=I$} and
\begin{equation}
\langle \Psi \vert S_{z} \vert \Psi \rangle =
2\alpha \frac{I}{I+1} S_{\text{intr}} \ .
\label{fb7-a} 
\end{equation}

\subsection{Estimate of the Schiff moment}
\label{subsec:Sciff-estimate-simplest}

Let us present a simple estimate of the nuclear Schiff moment in the
case there is a \mbox{T-,} P- odd interaction
and the nucleus possess a quadrupole and octupole deformation.
One needs to calculate the {\em intrinsic} moment and the mixing
coefficient $\alpha$. We consider an axially symmetric nucleus which
has a sharp surface and constant nucleon density $\rho_{\text{t}}$.
The surface of the nucleus has the form
\begin{equation}
R=R_{0} (1+ \beta_{1}Y_{10} +\beta_{2}Y_{20} +\beta_{3}Y_{30}) \ .
\label{surafce-est} 
\end{equation}
The $\beta_{1}$ deformation is introduced
\cite{Strutinsky-56,Dorso-Myers-Swiatecki,Bohr-Mottelson} 
to retain the center of mass at $z=0$:
\begin{equation}
\int z \rho_{\text{t}} d^{3}\bbox{r} 
\sim R_{0}(\sqrt{\frac{4\pi}{3}}\beta_{1}
+ \frac{9}{\sqrt{35}}\beta_{2}\beta_{3}) \ .
\label{centermass-est}
\end{equation}
Thus
\begin{equation}
\beta_{1}=-\sqrt{\frac{3}{4\pi}} \frac{9}{\sqrt{35}}
\beta_{2}\beta_{3} \ .
\label{beta1-est}
\end{equation}
 The proton density can be approximated by  constant
 \mbox{$\rho =3Z/4\pi R_{0}^{3}$}. The intrinsic electric    
dipole moment in this approximation is  zero, and only the first term in  
Eq.\ (\ref{fb5}) contributes to the intrinsic SM.
Using Eqs.\ (\ref{fb5},\ref{surafce-est},\ref{beta1-est}) one gets:
\begin{equation}
S_{\text{intr}}\equiv \frac{1}{10}e \int \rho r^{2} z d^{3}\bbox{r}=
e Z R_{0}^{3}\frac{9}{20\pi\sqrt{35}}\beta_{2}\beta_{3} \ .
\label{schiffmom-est}
\end{equation}
Using then the deformation parameters $\beta_{2}=0.12$
and $\beta_{3}=0.1$, $R_{0}=\mbox{$1.2\times A^{1/3}$}$~fm, 
\mbox{$A=230$}, and \mbox{$Z=88$} we obtain :
$S_{\text{intr}}=\mbox{$10.4 e$ fm$^3$}$.

The rotational states of an odd-mass nucleus
can be written in terms of intrinsic states as:
\begin{equation}
|I M \pm K \rangle =
\biggl(\frac{2I+1}{8 \pi^2}\biggr)^{1/2}
 D^I_{M \pm K} (\varphi,\theta,\psi) \varphi^{\text{(A)}}_{\pm K}
(\bbox{r}^\prime) \chi^{\text{(A)}}  \ ,
\label{ecn21a}
\end{equation}
where $D^I_{M \pm K}(\varphi,\theta,\psi)$ is a Wigner $D$-function,
$\chi^{\text{(A)}}$ is the wave function of the quadrupole and octupole
deformed (reflection asymmetric) nuclear core in the intrinsic frame,
and $\varphi^{\text{(A)}}_{\pm K}(\bbox{r}^\prime)$ is the wave
function of the unpaired nucleon in the intrinsic frame, with an
angular momentum projection of $\pm K$ on the $z'$ axis.

 Let us now present an order of magnitude estimate of
the mixing coefficient $\alpha$ which is needed to find the magnitude
of the Schiff moment in the laboratory frame (see Eq. (\ref{fb7-a})).
$\hat{K} = \bbox{I} \bbox{n}$ and $H_{TP}$ are both
T-, P-odd pseudoscalars. Therefore,
\mbox{$\langle \varphi^{\text{(A)}}_{+K} | H_{TP} |
 \varphi^{\text{(A)}}_{+K} \rangle \propto K$}  
and so 
$\mbox{$\langle \varphi^{\text{(A)}}_{-K} | H_{TP} |
  \varphi^{\text{(A)}}_{-K} \rangle$}=\mbox{$ 
-\langle \varphi^{\text{(A)}}_{+K} | H_{TP} |
\varphi^{\text{(A)}}_{+K} \rangle$}$  
(this fact can be easily supported by model calculations). Using this
fact and Eqs.\ (\ref{ecn21a}) and (\ref{fb6}) we get
$\mbox{$\langle \Psi^- | H_{TP} | \Psi^+ \rangle$}=\mbox{$
\langle \varphi^{\text{(A)}}_{+K} | H_{TP} | \varphi^{\text{(A)}}_{+K}
\rangle$}$. 
If single-particle wave function
$\varphi^{\text{(A)}}_{+K}$ was a good parity state this matrix
element would be zero. 
However, due to the perturbation caused by
the static octupole deformation of the nucleus ($V_3$),
it is a combination of the opposite parity spherical orbitals
$\phi_{1,+K}$ and $\phi_{2,+K}$ (for example, $s_{1/2}$ and $p_{1/2}$,
or $p_{3/2}$ and $d_{3/2}$, etc.):
\begin{equation}
\varphi^{\text{(A)}}_{+K} = \phi_{1,+K} + \gamma \phi_{2,+K} \ ,
\label{eae11a}
\end{equation}
\begin{equation}
\gamma = \frac{\langle \phi_{2,+K} | V_3 | \phi_{1,+K}
\rangle}{E_1 - E_2} \ ,
\label{eae11b}
\end{equation}
\begin{eqnarray}
\phi_{1,+K} & = & {\cal R}_1(r^{\prime})
\Omega_{j,l,+K}(\theta^{\prime}, \varphi^{\prime}) \ , 
\nonumber \\
\phi_{2,+K} & = & {\cal R}_2(r^{\prime}) \Omega_{j,\widetilde{l},+K}
(\theta^{\prime},\varphi^{\prime}) \nonumber \\
& = & - {\cal R}_2(r^{\prime}) (\bbox{\sigma} \hat{\bbox{r}}^{\prime}) 
\Omega_{j,l,+K} (\theta^{\prime},\varphi^{\prime}) \ ,
\label{eae11c}
\end{eqnarray}
where $\widetilde{l} = 2j-l$.
(Of course there will also be an admixture of
other opposite parity states
having different values of $j$. We neglect these states
for simplicity.) The mixing coefficient $\gamma$ is proportional
to the parameter of the octupole deformation $\beta_3$. 
Using Eq.
 (\ref{eae11b}) one can check that numerically $\gamma \simeq \beta_3$.
Therefore, we have
\begin{equation}
\alpha = \frac{\langle
\varphi^{\text{(A)}}_{+K}|H_{TP}|\varphi^{\text{(A)}}_{+K}
\rangle}{E^{+}-E^{-}} 
\simeq 2 \beta_3 \frac{\langle
\phi_{1,+K} | H_{TP} | \phi_{2,+K} \rangle}{E^{+}-E^{-}} \ .
\label{ena42}
\end{equation}

Finally, we must estimate the matrix element between spherical orbitals
\mbox{$\langle\phi_{1,+K} \vert H_{TP} \vert \phi_{2,+K} \rangle$}.
The T- and P- odd interaction between a
nonrelativistic unpaired nucleon and the nuclear core can be 
described
by the following effective Hamiltonian (see e.g.\
Refs.\ \cite{Sushkov-Flamb-Khripl,Khriplovich-book,Herczeg}):
\begin{equation}
H_{TP} = \eta \frac{G}{2 \sqrt{2} m} 
\bbox{\sigma} \bbox{\nabla} \rho_{t} \ ,
\label{ehpta}
\end{equation}
where $\bbox{\sigma}$ is twice the spin operator for this nucleon,
$\rho_{t}$ is the nucleon density of the nuclear core,
$G = \mbox{$1.0 \times 10^{-5}/{m}^2$}$ is the Fermi constant,
$m$ is the mass of the nucleon and $\eta$ is a dimensionless
constant that describes the strength of the interaction.

Using Eq.\ (\ref{ehpta}) for
the form of $H_{TP}$ and
$\rho_{t}(r^{\prime}) = 
\mbox{$\theta(r^{\prime}-R_0) 
/ ({\textstyle \frac{4}{3}} \pi {r_0}^3)$}$
(where $\theta$ is the step function) we get
\begin{equation}
H_{TP} = -\eta \frac{3G}{8 \pi \sqrt{2} m {r_0}^3} 
(\bbox{\sigma} \hat{\bbox{r}}^{\prime}) \delta(r^{\prime}-R_0) \ .
\label{eae31c}
\end{equation}
Here $r_0=1.2$~fm is the internucleon distance.
Using Eq.\ (\ref{eae11c}) and 
$\mbox{$(\bbox{\sigma} \hat{\bbox{r}}^{\prime})^2$} = 1$
gives
\begin{eqnarray}
&&\langle \phi_{1,+K} | H_{TP} | \phi_{2,+K} \rangle \nonumber \\
&&= \eta \frac{3G}{8 \pi \sqrt{2} m {r_0}^3}
{\cal R}_1(R_0) {\cal R}_2(R_0) {R_0}^2
\approx \frac{\eta}{A^{1/3}} 1 \mbox{ eV} \ ,
\label{eae31d}
\end{eqnarray}
where we have used $\mbox{${\cal R}_1(R_0) {\cal R}_2(R_0)$} \approx 
\mbox{$1.4/ {R_0}^3$}$ \cite{Bohr-Mottelson-1}.
Using $\mbox{$|E^{+}-E^{-}|$} = 50~\mbox{ keV}$,
$\beta_3 = 0.1$ (see data in the Tables),
and Eqs.\ (\ref{ena42}) and (\ref{eae31d})
gives (for $A=230$)
$|\alpha| \sim 
\mbox{$2 \beta_3 A^{-1/3} \eta~\text{eV} / |E^{+}-E^{-}|$}
\sim 7 \times 10^{-7} \eta$.
This provides the following estimate for the collective Schiff moment
in the laboratory frame:
\begin{eqnarray}
 S \sim \alpha S_{intr} &\sim& 
 0.05 e {\beta_2} {\beta_3}^2 Z A^{2/3} \eta
\,(r_0)^3~\text{eV} / |E^{+}-E^{-}| \nonumber \\
&\sim& 700 \times 10^{-8}~\eta~ e~\text{fm}^3 \ .
\label{eeome30}
\end{eqnarray}
We see that the collective Schiff moment is about 500 times larger
than the Schiff moment due to the unpaired nucleon in spherical
nucleus, $S \simeq \mbox{$1.5 \times 10^{-8} \eta e~\text{fm}^3$}$
\cite{Sushkov-Flamb-Khripl,FKS-86}. Note however the strong dependence
of the collective Schiff moment on the deformation parameters.

We should remark here that this is a schematic calculation and more
detailed and realistic calculations are presented in sections
\ref{sec:models} and \ref{sec:calsect}. Nevertheless the present
estimate represents the essence of this theory and the order of
magnitude estimate agrees with the detailed calculations.

\section{NUCLEAR MODELS OF THE T-, P-ODD MOMENTS}
\label{sec:models}

\subsection{Nuclear shape and intrinsic moments}
\label{subsec:core-shape-moments}

In this paper we consider the moments of heavy deformed nuclei in the
ground states. The main contribution to the electric moments comes
from the even-even core which is well described by the two-fluid
liquid drop model, 
see e.g.\ 
Refs.\ \cite{Myers-Swiatecki-Ann,Bohr-Mottelson-1,Bohr-Mottelson}.
The surface of an axially symmetric deformed nucleus is
\begin{equation}
R=c_{_V}(\beta)R_{0}\bigl (1+\sum_{l=1}\beta_{l}Y_{l0}\bigr ) \ ,
\label{surafce} 
\end{equation}
where 
\mbox{$\displaystyle 
c_{_V}=1-\frac{1}{\sqrt{4\pi}}\sum_{l=1}\beta_{l}^{2}$}
ensures the volume conservation and
\mbox{$R_{0}=r_{0}A^{1/3}$}. 
For the sake of brevity and because the nuclear deformations we deal
with are relatively small $\beta_{2}<0.2$, $\beta_{3,4}<0.1$
we will put in our discussion the coefficient 
\mbox{$c_{_V}(\beta)= 1$}. (In the actual calculations $R_{0}$ is
replaced by \mbox{$c_{_V}(\beta)R_{0}$}.)

If the nucleus is reflection-asymmetric then 
the $\beta_{1}$ deformation parameter 
is needed to keep the center of mass fixed at 
\mbox{$z=0$},
i.e.\ \mbox{$\int z d^{3}\bbox{r} =0$}.  
In lowest order in nuclear deformations 
\cite{Strutinsky-56,Dorso-Myers-Swiatecki,Bohr-Mottelson}
\begin{equation} 
\beta_{1}=-3\sqrt{\frac{3}{4\pi}}
\sum_{l=2}
\frac{(l+1) \beta_{l}\beta_{l+1}}{\sqrt{(2l+1)(2l+3)}} \ .
\label{beta-one}
\end{equation}
Due to the Coulomb force protons and neutrons are differently
distributed over the nuclear volume. From the requirement of a
minimum in the energy 
\cite{Myers-Swiatecki-Ann,Leander-Nazarewicz-Bertsch}
\begin{equation} 
\frac{\rho_{p}(\bbox{r})-\rho_{n}(\bbox{r})}
{\rho_{p}(\bbox{r})+\rho_{n}(\bbox{r})}=
-\frac{1}{4C} V_{\text{Coul}}(\bbox{r}) \ ,
\label{density-diff}
\end{equation}
where \mbox{$\rho_{p}\equiv \rho$} and $\rho_{n}$ are the proton and
neutron densities, $V_{\text{Coul}}(\bbox{r})$ is the Coulomb
potential created by $\rho_{p}(\bbox{r})$ and $C$ is the volume
symmetry-energy coefficient of the liquid-drop model.
To lowest order
\cite{Leander-Nazarewicz-Bertsch}
\begin{eqnarray}
\rho=&&\frac{\rho_{0}}{2} 
-\frac{\rho_{0}}{8} \frac{e^{2}Z}{CR_{0}}
\nonumber \\
&&\times \biggl [\frac{3}{2}-\frac{1}{2}\
\Bigl ( \frac{r}{R_{0}}\Bigr )^{2}
+\sum_{l=1} \frac{3}{2l+1}\Bigl (\frac{r}{R_{0}}\Bigr )^{l}
\beta_{l}Y_{l0}\biggr ] \ ,
\label{protdis}
\end{eqnarray}
where \mbox{$\rho_{0}=3A/(4\pi R_{0}^{3})$}.
The coefficient $C$ is not known very accurately, its value
for nuclei studied here is in the range \mbox{20--35 MeV} 
\cite{Leander-Nazarewicz-Bertsch,Dorso-Myers-Swiatecki,%
Butler-Nazarewicz}.  Note that requiring 
\mbox{$\int \rho(\bbox{r}) d^{3}\bbox{r} =Z$} 
one has in lowest order \cite{Myers-Swiatecki-Ann}
\begin{equation}
Z=\frac{1}{2}A \Bigl (1-\frac{3}{10}\frac{e^{2}Z}{CR_{0}} \Bigr ) \ .
\label{C-Z}
\end{equation}

We compute the intrinsic Schiff moment by substituting the density 
in Eq.\ (\ref{protdis}) into Eq.\ (\ref{fb5}). 
Because of the relative shift of protons versus neutrons 
the nucleus in the intrinsic frame has a dipole moment
as calculated in the past
\cite{Strutinsky-56,BM-57-59,Leander-Nazarewicz-Bertsch,%
Dorso-Myers-Swiatecki}
and given by
\begin{equation} 
d_{\text{intr}}=eAZ\frac{e^{2}}{C}\frac{3}{40\pi}\sum_{l=2} 
\frac{(l^{2}-1)(8l+9)}{[(2l+1)(2l+3)]^{3/2}}\beta_{l}\beta_{l+1} \ .
\label{dipmom}
\end{equation}
A more detailed treatment of the intrinsic dipole moment includes also
the neutron skin effect which reduces $d_{\text{intr}}$ somewhat
\cite{Dorso-Myers-Swiatecki,Butler-Nazarewicz}.
To discuss corrections to the intrinsic SM it is convenient to
decompose it into two terms 
\begin{equation}
S_{\text{intr}}=S^{(1)}_{\text{intr}} + S^{(2)}_{\text{intr}} \ .
\label{shiffmom-twoterms}
\end{equation}
The first term includes only
contribution from the constant part of the density, $\rho(r=0)$ in
Eq.\ (\ref{protdis}). 
In lowest order in deformation it equals to
\begin{equation}
S^{(1)}_{\text{intr}}=e AR_{0}^{3} \frac{3}{40\pi} \Bigl (
1-\frac{e^{2}Z}{R_{0}C}\frac{3}{8} \Bigr )
\sum_{l=2}
\frac{(l+1) \beta_{l}\beta_{l+1}}{\sqrt{(2l+1)(2l+3)}} \ .
\label{shiffmom-p1}
\end{equation}
As observed above this contribution comes from the first term in 
Eq.\ (\ref{fb5}) only. 
The second term is due to the Coulomb redistribution of the proton
density and stems from the last two terms in the brackets 
in Eq.\ (\ref{protdis}). A simple derivation gives
\begin{equation}
S^{(2)}_{\text{intr}}=e AR_{0}^{3} \frac{3}{40\pi}
\frac{e^{2}Z}{R_{0}C}\frac{29}{280}
\sum_{l=2}
\frac{(l+1) \beta_{l}\beta_{l+1}}{\sqrt{(2l+1)(2l+3)}} \ .
\label{shiffmom-p2}
\end{equation}
This term gives about 10\% contribution for nuclei with $Z\sim 90$.
Using Eq.\ (\ref{C-Z}) one can approximate the intrinsic SM as
\begin{equation}
S_{\text{intr}} \simeq e ZR_{0}^{3} \frac{3}{20\pi} 
\sum_{l=2}
\frac{(l+1) \beta_{l}\beta_{l+1}}{\sqrt{(2l+1)(2l+3)}} \ .
\label{shiffmom-propz}
\end{equation}
The contribution of the $\beta_{3}\beta_{4}$ term in both
$d_{\text{intr}}$ and $S_{\text{intr}}$, for the nuclei studied
here, is of about the same size as the contribution of the
$\beta_{2}\beta_{3}$ term.

The expression for the intrinsic octupole moment is
\cite{Bohr-Mottelson,Leander-Chen}
\begin{equation}
Q_{3~\text{intr}}=
\frac{3e ZR_{0}^{3}}{2\sqrt{7\pi}}(\beta_{3}+
\frac{2}{3}\sqrt{\frac{5}{\pi}}\beta_{2}\beta_{3} 
+\frac{15}{11\sqrt{\pi}}\beta_{3}\beta_{4} + ...) \ .
\label{oct-intr}
\end{equation}

Various nuclear surface corrections to the density of nucleons such as
the neutron skin are not included in the above equations 
for the SM. For the intrinsic dipole moment these corrections were 
included in Refs. \cite{Dorso-Myers-Swiatecki,Butler-Nazarewicz} 
using the droplet model. The corrections for $d_{\text{intr}}$ are 
of the same order as the main term in Eq.\ (\ref{dipmom}) but
have the opposite sign.  One can conclude from 
Eqs.\ (\ref{dipmom}-\ref{shiffmom-propz}) that since such corrections
alter only the term $S^{(2)}_{\text{intr}}$ they can contribute to
the Schiff moment at most 10\%.

Values of the Schiff moment obtained using 
Eq.\ (\ref{shiffmom-propz}) are about 30\% less than 
given by the direct calculation using Eqs.\ (\ref{fb5},\ref{protdis}) 
(with \mbox{$C\approx 27$ MeV} corresponding to the value obtained in
the droplet model which includes the effect of a neutron skin
\cite{Dorso-Myers-Swiatecki,Butler-Nazarewicz}).
This is partly due to terms of higher orders in the deformation not
included in Eqs.\ (\ref{C-Z}-\ref{shiffmom-propz}).

Both the intrinsic SM and dipole moments are 
second order in nuclear deformation and may turn out to be sensitive
to details of the proton density distribution.
Because of that it is important to take into account quantum
mechanical corrections to the liquid drop model. 
Such correction can be
included through the Strutinsky shell correction method
\cite{Strutinsky-shellcorr,Brack-Strutinsky-ea}.
In this method the level density is decomposed into a smooth averaged
density and a remaining part, fluctuating with the shell filling.
The corrected expectation value of an one-body operator, e.g.\ 
the intrinsic Schiff moment 
(for the similar treatment of the intrinsic dipole moment see
Refs.\ \cite{Leander-Nazarewicz-Bertsch,Butler-Nazarewicz})
is written as a sum of ``macroscopic'' and shell correction terms
\cite{Brack-Strutinsky-ea}
\begin{equation}
\tilde{S}_{\text{intr}}=S^{\text{M}}_{\text{intr}}+
S^{\text{shell}}_{\text{intr}} \ .
\label{shiffmom-macr-shell}
\end{equation}
As the ``macroscopic'' part one takes the liquid drop moment,
in the case of the Schiff moment $S^{\text{M}}_{\text{intr}}$ 
is given in lowest order in $\beta_{i}$ by 
Eq.\ (\ref{shiffmom-twoterms}) or 
by approximate Eq.\ (\ref{shiffmom-propz}).
The shell correction term is given by 
\cite{Brack-Strutinsky-ea,Leander-Nazarewicz-Bertsch,Butler-Nazarewicz}
\begin{equation}
S^{\text{shell}}_{\text{intr}}=\sum_{i} (v^{2}_{i}-n_{i}) 
\langle i\vert S \vert i\rangle \ ,
\label{shiffmom-shellcorr}
\end{equation}
where $v^{2}_{i}$ are the BCS quasiparticle occupation numbers and
$n_{i}$ - the smoothed single-particle occupation numbers, 
for the state $i$.
The latter determine the averaged level density. (Detailed
expressions can be found in 
Refs.\ \cite{Brack-Strutinsky-ea,Bolsterli-Nix-ea}.)

It is convenient to express the corrections relatively to the
$S^{\text{M}}_{\text{intr}}$.
The 
second term in Eq.\ (\ref{fb5})
is proportional to 
the intrinsic dipole moment
$d_{\text{intr}}$.
The shell correction to $d_{\text{intr}}$
was studied in detail in 
\cite{Leander-Nazarewicz-Bertsch,Butler-Nazarewicz}.
The results 
show that the shell correction to $d_{\text{intr}}$
is of the same order of magnitude as the ``macroscopic'' dipole moment
given by Eq.\ (\ref{dipmom}).
Therefore the subsequent correction to the second term of the
intrinsic SM in Eq.\ (\ref{fb5}) does not exceed (5--7)\% of the
value of $S^{\text{M}}_{\text{intr}}$.

The shell correction to the deviation of proton and
neutron centers of mass for octupole deformed actinide nuclei
was also investigated in the Refs.\
\cite{Leander-Nazarewicz-Bertsch,Butler-Nazarewicz}.
The deviation of the center of mass can be represented 
formally
as a change of the $\beta_{1}$ deformation parameter
(see Eq.\ (\ref{centermass-est})).
Note that the intrinsic Schiff moment is in the lowest order
proportional to the $\beta_{1}$ for protons 
(Eqs.\ (\ref{beta-one},\ref{shiffmom-p1}-\ref{shiffmom-propz})).
Therefore the resulting shell correction for the
first term of the intrinsic SM  in Eq.\ (\ref{fb5}) 
is analogous to the one obtained in 
\cite{Leander-Nazarewicz-Bertsch,Butler-Nazarewicz} for the
proton center of mass.

\subsection{Particle-core model for a reflection-asymmetric nucleus
and P-, T- odd mixing of parity doublets}
\label{subsec:particl-core-mod}

Any nucleus, whether even or odd mass that is reflection asymmetric
may possess {\em intrinsic} \mbox{T-,} P- odd moments.
Such moments will exist in the {\em laboratory} frame only if 
there is also a \mbox{T-,} P- odd mixing between g.s.\ parity doublet 
levels. If the even-even nucleus is perfectly axial- and spin
symmetric, a pseudoscalar \mbox{T-,} P- odd operator cannot mix
doublet states which have identical intrinsic structure \cite{SA-PLB}.
Besides, the energy splitting within parity doublets is 
systematically much less in odd (or odd-odd) than in
neighboring even-even nuclei \cite{Leander-82,Leander-Sheline,Brink}.
So in odd (or odd-odd) nuclei the mixing should be considerably
larger. For this reason we consider odd nuclei in what follows.

We use here the particle-core model for a reflection-asymmetric
nucleus \cite{Leander-Sheline,Brink}. The \mbox{T-,} P- odd as well
as P- odd T- even mixing was studied in this model recently
\cite{ABS-doublets,SA-PLB}.  
This model involves two \mbox{$K^{\pi}=0^{+}$} and
\mbox{$K^{\pi}=0^{-}$} members of a parity doublet of the even-even
core with the energy splitting $E_{c}$ between them.  The wave
functions of these states $\chi^{\pi}$ are projections of the
reflection-asymmetric core state $\chi_{_{A}}$ \cite{Leander-Sheline}
\begin{equation}
\chi^{\pi}=\frac{1}{\sqrt{2}}(1+\pi \hat{P})\chi_{_{A}} \ ,
\end{equation}
where $\hat{P}$ is the core parity operator.
The single-particle states $\varphi_{_{A}}$
are solutions of the reflection-asymmetric single-particle plus
pairing Hamiltonian \mbox{$H_{\text{s-p}}+H_{\text{pair}}$}
(different forms of potentials such as
folded Yukawa, deformed Woods-Saxon or Nilsson are used 
in $H_{\text{s-p}}$
\cite{Leander-Sheline,Leander-Chen,WSBETA,SA-PLB}).
The wave functions in the model are \cite{Leander-Sheline,SA-PLB}:
\begin{equation}
\Psi^{Ip}_{MK}=\biggl [\frac{2I+1}{16\pi^{2}}\biggr ]^{1/2}
[1+\hat{R}_{2}(\pi)]D^{I}_{MK}
{\Phi}_{K}^{p} \ ,
\label{14h}
\end{equation}
where \mbox{$\hat{R}_{2}(\pi)$} denotes rotation through an angle 
$\pi$ about the intrinsic 2 axis.
The \mbox{${\Phi}^{p}\equiv {\Phi}^{\pm}$} 
are particle-core intrinsic states of good parity $p$. 
Denoting the good parity particle states ${\phi}^{\pi}$
we write (in the matrix notation 
$a{\phi}=\mbox{$\sum_{k} a_{k}{\phi}_{k}$}$)
\begin{eqnarray}
&&{\Phi}^{+}=a_{+}{\chi}^{+}{\phi}^{+} + 
b_{+}{\chi}^{-}{\phi}^{-}
\nonumber \\ 
&&{\Phi}^{-}=a_{-}{\chi}^{-}{\phi}^{+} +
b_{-}{\chi}^{+}{\phi}^{-} \ .
\label{12h}
\end{eqnarray}
The matrix elements of $V^{\text{PT}}$ are given by \cite{SA-PLB}
\begin{eqnarray}
&&\langle \Psi^{I+}_{MK}\vert V^{\text{PT}} 
\vert \Psi^{I-}_{MK}\rangle
\nonumber \\
&&=a_{+}b_{-}
\langle \phi^{+}_{K}\vert V^{\text{PT}} \vert \phi^{-}_{K}\rangle
+a_{-}b_{+}
\langle \phi^{+}_{K}\vert V^{\text{PT}} \vert \phi^{-}_{K}\rangle \ .
\label{22ah}
\end{eqnarray}
(As already mentioned the pseudoscalar operator $V^{\text{PT}}$ cannot
connect states of an even-even spin symmetric and axially symmetric
core $\chi^{\pi}$ \cite{SA-PLB}.)

The expectation value of a \mbox{T-,} P- odd operator $\hat{O}$ in
a \mbox{T-,} P- admixed state ${\tilde{\Phi}}_{i}^{+}$ is
\begin{equation}
\langle {\tilde{\Phi}}_{i}^{+} \vert \hat{O} \vert
{\tilde{\Phi}}_{i}^{+}\rangle =
2\alpha_{ii} \langle {\Phi}_{i}^{+}\vert \hat{O} \vert
{\Phi}_{i}^{-} \rangle  +
2\sum_{j\neq i}\alpha_{ij}
\langle {\Phi}_{i}^{+}\vert \hat{O} \vert
{\Phi}_{j}^{-} \rangle  \ .
\label{exp-value}
\end{equation}
The matrix elements between core states are
\begin{equation}
\langle \chi^{+} \vert \hat{O} \vert \chi^{-} \rangle=
\langle \chi_{_{A}} \vert \hat{O} \vert \chi_{_{A}} \rangle \ .
\label{mom-core}
\end{equation}
One can write the one-body operator $\hat{O}$ as the sum of core and
particle terms
\mbox{$\hat{O}=\hat{O}_{\text{core}}+\hat{O}_{\text{p}}$} and obtain
\begin{eqnarray}
&&\langle {\Phi}_{i}^{+}\vert \hat{O} \vert
{\Phi}_{j}^{-} \rangle=
\langle \chi_{_{A}} \vert \hat{O}_{\text{core}} \vert 
\chi_{_{A}} \rangle (a_{{+}~{i}}a_{{-}~{j}}+b_{{+}~{i}}b_{{-}~{j}})
\nonumber \\
&&
+\langle {\phi}^{+}_{i} \vert \hat{O}_{\text{p}} \vert 
{\phi}^{-}_{j} \rangle 
a_{{+}~{i}}b_{{-}~{j}}+
\langle {\phi}^{+}_{j} \vert \hat{O}_{\text{p}} \vert
{\phi}^{-}_{i} \rangle 
a_{{-}~{j}}b_{{+}~{i}} \ .
\label{core-plus-part-mom}
\end{eqnarray}
For \mbox{$i=j$} this is just the intrinsic moment
in the reflection-asymmetric core-particle state
${\Phi}=\mbox{$\chi_{_{A}}\varphi_{_{A}}$}\equiv
\mbox{$\chi_{_{A}}(a\phi^{+} + b\phi^{-})$}$.
For closely spaced doublets 
\mbox{$a_{+~i}\approx a_{-~i}$}, \mbox{$b_{+~i}\approx b_{-~i}$}
and
\mbox{$(a_{{+}~{i}}a_{{-}~{j}}+b_{{+}~{i}}b_{{-}~{j}})\approx
\delta_{ij}$}.
With respect to the single-particle contribution to 
a \mbox{T-,} P- odd moment 
in Eqs.\ (\ref{exp-value},\ref{core-plus-part-mom}) note that
the admixture of the doublet level is considerably
larger than the admixtures of the other levels of opposite parity 
i.e.\ \mbox{$\alpha_{ii}\gg \alpha_{ij}, i\neq j$}.
Neglecting these ``off-diagonal'' $\alpha_{ij}$ 
contributions  one gets 
\begin{mathletters}
\label{mom-expvalues}
\begin{equation}
\langle {\tilde{\Phi}}_{i}^{+} \vert \hat{O} \vert 
{\tilde{\Phi}}_{i}^{+}\rangle \approx
2\alpha_{ii}\Bigl (
\langle \chi_{_{A}} \vert \hat{O}_{\text{core}} \vert 
\chi_{_{A}} \rangle  + 
\langle \varphi_{_{A}}  \vert \hat{O}_{\text{p}} \vert 
\varphi_{_{A}} \rangle \Bigr ) \ ,
\label{mom-expvalues-a}
\end{equation}
and 
\begin{eqnarray}
&&\langle \tilde{\Psi}^{I+}_{MK}\vert\hat{O}_{\mu=0}\vert
\tilde{\Psi}^{I+}_{MK} \rangle \nonumber \\
&&=\langle IM l0 \vert IM \rangle
\langle IK l0 \vert IK \rangle
\langle {\tilde{\Phi}}_{i}^{+} \vert \hat{O}_{\mu=0} \vert 
{\tilde{\Phi}}_{i}^{+}\rangle \ ,
\label{mom-expvalues-b}
\end{eqnarray}
\end{mathletters}
where $l$ is the rank of the operator.
For a vector operator such as $\hat{\bbox{S}}$ (or $\hat{\bbox{d}}$)
one obtains in accordance with
Eqs.\ (\ref{simple-schiff-dipole},\ref{fb7},\ref{fb7-a}) 
\begin{eqnarray}
&& \langle \tilde{\Psi}^{I+}_{MK}\vert\hat{S}_{z}\vert
\tilde{\Psi}^{I+}_{MK} \rangle \nonumber \\
&&=\frac{MK}{I(I+1)}
2\alpha_{ii}\Bigl (
\langle \chi_{_{A}} \vert \hat{S}_{\text{core}} \vert 
\chi_{_{A}} \rangle  + 
\langle \varphi_{_{A}}  \vert \hat{S}_{\text{p}} \vert 
\varphi_{_{A}} \rangle \Bigr ) \ .
\label{mom-expvalues-vector}
\end{eqnarray}
The intrinsic SM of the even $Z$ core 
\mbox{$\langle \chi_{_{A}} \vert \hat{S}_{\text{core}} \vert 
\chi_{_{A}} \rangle$} is given by Eq.\ (\ref{shiffmom-macr-shell}).

We stress again that the essential difference between the
reflection-asymmetric (octupole deformed) nucleus and the reflection
symmetric (quadrupole deformed) one is that the former has
{\em intrinsic} T-, P- odd moments which are essentially
{\em collective}
since they involve contributions of the core nucleons.
Note that in a nucleus which has a quadrupole deformation but 
no octupole deformation the ground state and its parity ``partner''
are built from a single good parity core state, e.g.\ $\chi^{+}$, 
and therefore the coefficients in Eq.\ (\ref{12h}) are then 
$a_{+}=1$, $b_{+}=0$,  $a_{-}=0$, $b_{-}=1$. 
So apart from a small core polarization contribution 
\cite{Sushkov-Flamb-Khripl,FKS-86} 
there is no contribution from the core.

The parity mixing of a reflection asymmetric single particle state
$\varphi_{_{A}}=\mbox{$\sum_{k} a_{k}\phi^{+}_{k}+
\sum_{m} b_{m}\phi^{-}_{m}$}$ 
is conveniently expressed via
the expectation value of the single-particle parity operator
$\hat{\pi}_{\text{p}}$ \cite{Leander-Sheline}:
\begin{equation}
\pi_{\text{p}}\equiv\langle \varphi_{_{A}}\vert \hat{\pi}_{\text{p}}
\vert \varphi_{_{A}}\rangle =\sum_{k} a_{k}^{2}-\sum_{m} b_{m}^{2} \ .
\label{parity-sp}
\end{equation}
Using this quantity one can write the admixture coefficient
$\alpha_{ii}$ as 
\begin{equation}
\vert \alpha_{ii} \vert \approx
\Bigl \vert
\frac{\langle \phi^{+}_{i}\vert V^{\text{PT}}\vert\phi^{-}_{i}\rangle}
{E_{i}^{^+}-E_{i}^{^-}} \Bigr \vert 
\sqrt{(1-\pi_{\text{p}i}^{2})} \ .
\label{alpha-ii-approx}
\end{equation}
Here $E_{i}^{^+}$ and $E_{i}^{^-}$ are the energies of the
particle-core states ${\Phi}_{i}^{+}$, ${\Phi}_{i}^{-}$.
The $\alpha_{ii}$ obviously is maximal for a doublet built
on the strongly parity admixed 
\mbox{$(\vert\pi_{\text{p}}\vert \ll 1)$} intrinsic state.
Note, that although the form of Eq.\ (\ref{alpha-ii-approx}) is 
analogous (for \mbox{$\pi_{\text{p}i}=0$}) to the case
when there
is a close single-particle level of opposite parity
in reflection-symmetric deformed nuclei \cite{Sushkov-Flamb-Khripl},
there is an important difference.
Namely \cite{Sushkov-Flamb-Khripl}, if one neglects the spin-orbit
interaction and assumes that the nuclear density
\mbox{$\rho_{\text{t}}(\bbox{r})$} and the single-particle potential
\mbox{$U(\bbox{r})$} have the same form then
\begin{equation}
V^{\text{PT}} \sim \bbox{\sigma\nabla}U(\bbox{r})=
i [\bbox{\sigma p},H_{\text{s-p}}] \ ,
\label{vpt-approx}
\end{equation}
and the one body matrix element 
\mbox{$\langle \phi^{+}\vert V^{\text{PT}}\vert\phi^{-}\rangle$}
is proportional to the energy difference
\begin{equation}
\langle \phi^{+}\vert V^{\text{PT}}\vert\phi^{-}\rangle \sim
i \langle \phi^{+}\vert [\bbox{\sigma p},H_{\text{s-p}}] 
\vert\phi^{-}\rangle \sim e_{{\phi^{-}}}-e_{{\phi^{+}}} \ ,
\label{refl-symm-vpt}
\end{equation}
where $e_{{\phi^{+}}}$ and $e_{{\phi^{-}}}$ are the single-particle
energies. 
In the case of a reflection symmetric (e.g.\ quadrupole deformed)
nucleus
\mbox{$E^{^+}=e_{{\phi^{+}}}$} and \mbox{$E^{^-}=e_{{\phi^{-}}}$}.
The mixing coefficient is therefore enhanced only in a few
cases when single-particle deformed levels are accidentally close
and the approximation of 
Eqs.\ (\ref{vpt-approx},\ref{refl-symm-vpt}) becomes a crude one 
\cite{Sushkov-Flamb-Khripl}.
In the reflection-asymmetric case always 
\cite{Leander-Sheline,Brink,SA-PLB}
\begin{equation}
\vert E^{^+}-E^{^-}\vert \lesssim
\frac{E_{c} \vert e_{{\phi^{+}}}-e_{{\phi^{-}}} \vert}
{2\vert\langle\phi^{+}\vert V_{\text{odd}}\vert \phi^{-}\rangle\vert}
\ll \vert e_{{\phi^{+}}}-e_{{\phi^{-}}} \vert \ ,
\label{energ-diff}
\end{equation}
where $V_{\text{odd}}$ is the reflection-asymmetric
part of the single-particle Hamiltonian $H_{\text{s-p}}$.
Therefore the admixture coefficient $\alpha_{ii}$
is generally enhanced relatively to 
that in spherical or quadrupole deformed nuclei.
The splitting within the parity doublet 
(for \mbox{$K\neq\frac{1}{2}$}) can be also written as 
\cite{Leander-Sheline,Brink} 
\begin{equation}
\vert E^{^+}-E^{^-}\vert \simeq\vert
E_{c}\pi_{\text{p}}\vert \ , 
\label{sp-core-splitt}
\end{equation}
exhibiting the reduction of splitting for an odd nucleus
in comparison with the splitting of
parity doublet states of the even-even core.

Let us look now at the single-particle part of the SM given by 
Eqs.\ (\ref{fb5},\ref{core-plus-part-mom},\ref{mom-expvalues}).
The ``diagonal'' single-particle contribution 
in Eq.\ (\ref{mom-expvalues-a}) can be written as 
\begin{eqnarray}
&&2\alpha_{ii} 
\langle {\phi}^{+}_{i} \vert \hat{O}_{\text{p}} \vert 
{\phi}^{-}_{i} \rangle 
(a_{{+}~{i}}b_{{-}~{i}}+a_{{-}~{i}}b_{{+}~{i}}) \nonumber \\
&&\approx 2 
\langle {\phi}^{+}_{i} \vert \hat{O}_{\text{p}} \vert
{\phi}^{-}_{i} \rangle
\frac{\langle \phi^{+}_{i}\vert V^{\text{PT}}\vert\phi^{-}_{i}\rangle}
{E_{i}^{^+}-E_{i}^{^-}}  (1-\pi_{\text{p}i}^{2}) \ .
\label{sp-approx}
\end{eqnarray}
Thus, in general, due to the enhancement in $\alpha_{ii}$ the
single-particle contribution of the odd proton is enhanced relatively
to the case of a quadrupole deformed but reflection symmetric
nucleus.  In the proton-odd nuclei the single proton contributes to
both terms of the SM operator in Eq.\ (\ref{fb5}).  These terms are
of the same order and partially cancel.  In the neutron-odd nuclei
the single neutron contribution to the SM operator in 
Eq.\ (\ref{fb5}) enters only through the corrections to the dipole
moment term and therefore is very small.

As is shown in Ref.\ \cite{Leander-Sheline} the effects of BCS-like
pairing are maximal when the Fermi level is exactly halfway between
the single-particle levels. In this case the simplest model of two
opposite parity single-particle levels results in an exactly
degenerate parity doublet.  In realistic calculations the effect of
pairing correlations on single-particle levels is revealed only in
somewhat altered energy splittings between members of parity doublets.
In the calculations of matrix elements of parity and parity and time
reversal violating potentials and single-particle parts of electric
moments, the effect of pairing is more important. 
When one uses the BCS quasiparticle states instead of single-particle
states, the matrix elements of the
operators $V^{\text{PT}}$, $V^{\text{PV}}$ and $\hat{O}_{\text{p}}$
are to be multiplied by pairing factors 
\mbox{$u_{f}u_{i}+c_{\text{sym}}v_{f}v_{i}$}, 
where $c_{\text{sym}}$ depends on the symmetry properties 
of the operator under the time reversal
and Hermitian conjugation \cite{Bohr-Mottelson,Bohr-Mottelson-1}.
For Hermitian operators, \mbox{$c_{\text{sym}}=1$} if an operator
is T-odd and \mbox{$c_{\text{sym}}=-1$} if it is T-even 
\cite{Bohr-Mottelson-1}. 
Thus, for the T-odd $V^{\text{PT}}$, one has
the factor \mbox{$u_{f}u_{i}+v_{f}v_{i}$},
whereas for the T- even $V^{\text{PV}}$ and a single-particle 
matrix element of an electric operator $\hat{O}_{\text{p}}$
the corresponding factor is
\mbox{$u_{f}u_{i}-v_{f}v_{i}$}.
The difference in signs in the
expressions for pairing correlation factors is related to the fact
that matrix element of $V^{\text{PV}}$ operator between members of a
parity doublet goes to zero when the doublet becomes degenerate
while the matrix element of $V^{\text{PT}}$ operator does not vanish
\cite{Sushkov-Flamb-f,SA-PLB,Desplanques-Noguera-review}.

\subsection{The P- and T- odd interaction}
\label{subsec:pt-odd-mixing}

The P- and T- odd nucleon-nucleon two-body potential can be written
in the form \cite{FKS-86,Khriplovich-book} as
\begin{eqnarray}
W_{ab}&&=\frac{G}{\sqrt{2}}\frac{1}{2m}\Bigl (
(\eta_{ab}\bbox{\sigma}_{a} - \eta_{ba}\bbox{\sigma}_b)
\bbox{\nabla}\delta_{a}(\bbox{r}_a -\bbox{r}_b)
\nonumber \\
&&
+
\eta'_{ab} [\bbox{\sigma}_a \times \bbox{\sigma}_b]
\{(\bbox{p}_a -\bbox{p}_b),\delta(\bbox{r}_a - \bbox{r}_b)\}\Bigr ) \ ,
\label{vpt-twobody}
\end{eqnarray}
where \mbox{$G=10^{-5}/m^{2}$} is the Fermi constant 
($m$ is the nucleon mass),
\mbox{$a,b$} designate a proton or neutron
and the curly brackets denote the anticommutation operation.
Note, that $\eta_{ab}$ are in fact effective constants.
In this work we use the corresponding effective
one-body P- and T- odd potential
\cite{Sushkov-Flamb-Khripl,Khriplovich-book,Herczeg}
\begin{equation}
V^{\text{PT}}=\frac{G}{\sqrt{2}}\frac{\eta}{2m}\rho_{0}
\sum_{i}
\bbox{\sigma}_{i}
(\bbox{\nabla}_{i}f(\bbox{r}_{i})) \ ,
\label{23h}
\end{equation}
where \mbox{$\rho_{\text{t}}(\bbox{r})=\rho_{0} f(\bbox{r})$}
is the nuclear density.
We assume \cite{FKS-86,Khriplovich-book}
that effective constants $\eta_{ab}$ and $\eta$ are of the same order
in magnitude. For the deformed nuclear density we use the expansion
\cite{Bohr-Mottelson} 
\widetext
\begin{eqnarray}
\rho(\bbox{r})= 
\rho_{0}(r)&-&R_{0}\frac{\partial\rho_{0}}{\partial r}
\biggl (\sum\beta_{l} Y_{l0}-\frac{c}{4\pi}\sum\beta^{2}_{l}+
\frac{1}{2}(r-R_{0})R_{0}\Bigl (\sum\beta_{l}
{\bbox\nabla}Y_{l0}\Bigr )^{2}\biggr )
\nonumber \\
&+&\frac{1}{2}R_{0}^{2}\frac{\partial^{2}\rho_{0}}{\partial r^{2}}
\Bigl (\sum\beta_{l} Y_{l0}\Bigr )^{2} + ... \ ,
\label{rho-expansion}
\end{eqnarray}
\narrowtext
where $\rho_{0}(r)$ has the usual Woods-Saxon form,
$\beta_{i}$ are the nuclear deformation parameters 
in Eq.\ (\ref{surafce}) and
\mbox{$\displaystyle
c=\frac{R_{0}}{2} \frac{\int\rho_{0}dr}{\int \rho_{0}r dr}$}.

\section{CALCULATIONS AND ESTIMATES}
\label{sec:calsect}

\subsection{Nuclear structure calculations}
\label{subsec:calc-nuclear}

The set of deformation parameters $\beta_{l}$ (up to $l=6$) 
was taken from  Ref.\ \cite{Cwiok-Nazarewicz}
where it was calculated using a deformed
Woods-Saxon single-particle potential and
Strutinsky shell correction method.
  From the sets of deformations, calculated in
Ref.\ \cite{Cwiok-Nazarewicz}, we have chosen those
which correspond to the experimental value of $K$ in the g.s.
This set is shown in Table \ref{deformations-table}.
The particle-core wave functions were
calculated using 
the reflection-asymmetric particle core model \cite{Leander-Sheline}.
We calculated both the intrinsic \mbox{T-}, P- odd moments and
admixture coefficients using basically the same single-particle
potential and
shell correction parameters.
The computer code WSBETA \cite{WSBETA} and the ``universal'' set of
parameters was used in calculations involving deformed single-particle
Woods-Saxon potential.
(A modification of the code was made to include explicitly
the $\beta_{1}$ deformation.)

To check the stability of the values of admixture coefficients we
calculated them also using a Nilsson potential with $\epsilon_{l}$
deformations approximately corresponding to the same nuclear surface
(see e.g.\ Ref.\ \cite{Moller-tabl95}).  
It is known \cite{Bengtsson-phys-scripta} that there are some
differences in the energies and wave functions when calculated with
deformed Nilsson or Woods-Saxon potentials in the actinide region.
Especially the proton levels are different because
of the fact that the Coulomb term is included in the Woods-Saxon
potential but is only simulated in the Nilsson potential.
Because of that we performed calculation of $\alpha$ - admixtures with
Nilsson potential only for the neutron-odd nuclei.
For the Nilsson potential we used parameters 
of Ref.\ \cite{Bengtsson-phys-scripta}
which are known to produce a good fit to experimental data. 
See for example the reflection-asymmetric calculations of $^{225}$Ra
in Ref.\ \cite{Sheline-Ragnarsson}.

The energy splittings of the core parity doublets - $E_{c}$
and moments of inertia were taken from Ref.\ \cite{Leander-Chen}.
The values of $E_{c}$ we used are given in Table
\ref{deformations-table}.
The effect of the 
\mbox{$l=1$} deformation 
($\beta_{1}$ or $\epsilon_{1}$) in the deformed single-particle
potential on the energies and admixture coefficients of 
doublets is small. However this
deformation is important 
for the calculation of the shell correction to the
intrinsic Schiff moment. 

For the BCS treatment of pairing we used the strength parameter
$G_{\text{p,n}}$ \cite{Nilsson69}
\begin{equation}
G_{\text{p,n}}=\frac{1}{A}\Bigl (g_{0}\pm g_{1}\frac{N-Z}{A}\Bigr ) \ ,
\label{pairing-strength}
\end{equation}
with \mbox{$g_{0}=19.2$ MeV}, \mbox{$g_{1}=7.4$ MeV} for
neutrons and \mbox{$g_{0}=22.0$ MeV}, \mbox{$g_{1}=8.0$ MeV} for
protons. 
We found that the admixture coefficients,
as well as other properties of these nuclei calculated
in Ref.\ \cite{Leander-Chen}, are not sensitive 
to the pairing strength (or gap parameter $\Delta$).
The standard Strutinsky shell correction calculation of the smoothed
single-particle occupation numbers $n$ was performed using parameters
taken from Ref.\ \cite{Bolsterli-Nix-ea}.

The intrinsic SM was calculated using 
Eqs.\ (\ref{shiffmom-macr-shell},\ref{shiffmom-shellcorr}). 
The ``macroscopic''
term $S^{\text{M}}_{\text{intr}}$ was computed directly using 
Eqs.\ (\ref{fb5},\ref{protdis}) with \mbox{$C=27$ MeV} 
which is the droplet model value \cite{Butler-Nazarewicz}.
We calculated the shell correction for the first term of the intrinsic
SM operator in Eq.\ (\ref{fb5}). 
This correction strongly depends on the proton
number and for \mbox{86$\leq$ Z$\leq 91$} its absolute value decreases
with increasing of Z. 
For Z$<$92 it has a negative
sign relatively to $S^{\text{M}}_{\text{intr}}$ and  
changes from 33\% for $^{223}_{~86}$Rn to 9\% for $^{225}_{~89}$Ac.
In case of proton-odd nuclei we calculated also the single proton
contribution to the SM. For the g.s.\ of $^{223}$Fr, $^{225}$Ac and
$^{229}$Pa it amounts to 2--5\% of the corresponding values of
$S^{\text{M}}_{\text{intr}}$. The resulting intrinsic Schiff moments of
octupole deformed nuclei are in the range
\mbox{(15--28)$e$~fm$^3$}. 
We estimate that when we allow for reasonable changes in the
parameters used, and when other corrections (not treated here)
are introduced the {\em intrinsic} SM will change by 30\% at most.
Thus we believe, that the uncertainties in the evaluated values of the
{\em intrinsic} SM are of the same order.

The computed values of {\em intrinsic} Schiff moments, admixture
coefficients, 
and resulting  Schiff moments in the laboratory system
as well as the calculated and
experimental energy splitting for the g.s.\ parity doublets and
calculated parities $\pi_{\text{p}}$
of the intrinsic single-particle g.s.\
are all given in Table \ref{moments-table-big}.  

The main uncertainty in the entire calculation of Schiff moments 
in the laboratory frame arises from the estimate of
admixture coefficients $\alpha$, which are
calculated using theoretical values of the energy splitting
between members of the doublets.
In our calculations the first state above the Fermi level
which has the same value of $K$ as the experimentally determined g.s.\
was chosen to be the g.s.\ level.
In the work of Leander and Chen \cite{Leander-Chen} the nonadiabatic
Coriolis coupling and other refinements were introduced 
(in some cases also adjustments of quasiparticle energies were made),
which allowed to describe properties of g.s.\ and excited levels.
We did not include couplings between states with different $K$ and
did not adjust model parameters to fit properties of individual states.
In some cases the states calculated as the lowest ones above the Fermi
surface do not have the experimentally determined values of $K$
\cite{Cwiok-Nazarewicz}.
For example the \mbox{$I=\frac{7}{2}$} g.s. of $^{223}$Rn 
which was described in \cite{Leander-Chen} as arising 
from Coriolis coupling of $K=\frac{7}{2}$ and $K=\frac{1}{2}$ states,
is not the lowest state 
in our calculation but is 270~keV above the g.s.
Thus it cannot be ruled out that in some cases some single-particle
states close to the Fermi level and different from the ones we chose,
fit better the g.s.\ parity doublet.
The absolute values of the admixture coefficients for the doublet
which was taken as the g.s.\ are 
for all nuclei we calculated 
(except $^{229}$Pa which has an exceptionally small energy splitting  
in the g.s.\ parity doublet)
in the range  \mbox{(1--5)$\times 10^{-7} \eta$} 
for both reflection asymmetric
Woods-Saxon and Nilsson single-particle potentials.
If we consider also admixture coefficients for the
two parity doublets closest to the g.s.\ parity doublet 
which have the same $K$ the range is 
\mbox{(0.15--5)$\times 10^{-7} \eta$}.
We remark that because the one-body P-, T- odd potential 
$V^{\text{PT}}$ in Eq.\ (\ref{23h}) is proportional to the
derivative of the density of nucleons and thus is surface-peaked,
its matrix elements depend on the behavior of the
single particle wave functions in the nuclear surface region and
hence may vary significantly from level to level.

As one sees in Table \ref{moments-table-big}
some of parities $\pi_{\text{p}}$ depart considerably from 
\mbox{$\pm 1$} meaning that these orbitals are
strongly parity-mixed, which is one of
the reasons for large admixture coefficients $\alpha$.
Eq.\ (\ref{sp-core-splitt}) describes well the
splittings within the doublets except for the 
\mbox{$K=\frac{1}{2}$} cases of the $^{225}$Ra and $^{221}$Fr g.s.,
where there is an additional Coriolis splitting 
\cite{Leander-Sheline,SA-PLB}.
For $^{225}$Ra, $^{221}$Fr and $^{225}$Ac the experimental energy
splitting between members of the g.s.\ parity doublet is well
reproduced in our calculation whereas in $^{223}$Fr and $^{223}$Ra
nuclei it differs by a factor of 2 or 3.
For $^{223}$Rn no data on doublets are available.
We should remark here that in view of 
Eq.\ (\ref{refl-symm-vpt})
one can expect some correlation between the matrix element of
$V^{\text{PT}}$ and the energy splitting within the doublet.
We expect therefore that the use of experimental energy splittings
will not necessarily lead to more precise values for the admixture
coefficients $\alpha$.

The nuclear Schiff moments for the octupole deformed nuclei
calculated here are about two orders of magnitude larger that those
obtained in Refs.\ \cite{Sushkov-Flamb-Khripl,FKS-86} for
isotopes of $^{129,131}$Xe, $^{199,201}$Hg and $^{203,205}$Tl.
As already mentioned the enhancements we discuss are for nuclei with
asymmetric shapes and therefore there is also two orders of magnitude
enhancement with respect to nuclei such as $^{161}$Dy and $^{237}$Np
\cite{Sushkov-Flamb-Khripl} which have large quadrupole deformation.
These nuclei have close to the g.s.\ levels of the same spin and
opposite parity as the g.s., but are believed to be
reflection symmetric. 

The SMs in even $Z$ nuclei such as Xe and Hg are caused mainly by
the polarization of protons in the core by the \mbox{T-}, P- odd field
of the external nucleon (see Eq.\ (\ref{vpt-twobody})).
Therefore, the SM is proportional to the $\eta_{np}$ constant. 
It was demonstrated in  \cite{FKS-86} that this ``polarization''
mechanism gives SM of the same order of magnitude as in nuclei that
have a proton outside the core.
Many-body corrections give also contributions proportional to other
\mbox{T-}, P- odd constants, e.g.\ $\eta_{nn}$. 
In our calculations the interaction of the odd proton or neutron 
with the even-even core and the nuclear SM in the laboratory system
are expressed via the effective constant $\eta$.

The experimental data regarding the nuclei we consider are
discussed in the reviews 
\cite{Leander-Chen,Jain-Sheline-RMP,Ahmad-Butler,%
Butler-Nazarewicz-RMP}. The $^{223}$Ra and $^{225}$Ra nuclei were
also considered in detail in Refs. 
\cite{Sheline-Chen-Leander-ra223,Sheline-Ragnarsson}.
In both $^{221,223}$Rn isotopes the \mbox{$I=\frac{7}{2}$} g.s.\
spin was determined using laser spectroscopy methods
\cite{laser-rn}, and no data on excited levels are presently
available. The computed deformations in these two isotopes calculated
in Ref.\ \cite{Cwiok-Nazarewicz} are similar. 
Using the Coriolis coupling
Leander and Chen \cite{Leander-Chen} were able to reproduce
spectroscopic characteristics of g.s.\ somewhat better 
for $^{223}$Rn than for $^{221}$Rn. 
We made a calculation for $^{223}$Rn because its
\mbox{$I=\frac{7}{2}$} g.s.\ is easier to interpret theoretically
\cite{Cwiok-Nazarewicz,Leander-Chen}. 

The case of Fr is especially interesting 
in light of the recent experiments involving trapping of Fr atoms
\cite{trapping-Fr}.
The latest experimental and theoretical studies of 
$^{221}$Fr \cite{Sheline-221Fr-90} 
and $^{223}$Fr 
\cite{Kurcewicz-223Fr,Sheline-223Fr}
provide strong evidence of intrinsic reflection asymmetry in the g.s.
The g.s of $^{221}$Fr is \mbox{$K^{\pi}=\frac{1}{2}^{-}$},
\mbox{$I^{\pi}=\frac{5}{2}^{-}$}.
The assignment 
of the doublet {$K^{\pi}=\frac{1}{2}^{+}$},
\mbox{$I^{\pi}=\frac{5}{2}^{+}$} state at 234~keV 
was suggested in \cite{Sheline-221Fr-90}.
The result for the Schiff moment of $^{221}$Fr is smaller than that of
$^{223}$Fr because of the following factors. Firstly, the
factor $\frac{MK}{I(I+1)}$ in Eq.\ (\ref{mom-expvalues-vector})
is $\frac{1}{7}$ for $^{221}$Fr, whereas for  
\mbox{$I=\frac{3}{2}$}, \mbox{$K=\frac{3}{2}$} g.s.\ of $^{223}$Fr
it is $\frac{3}{5}$. Secondly our calculation gives the admixture
coefficient $\alpha$ for $^{221}$Fr about 3 times less than for 
$^{223}$Fr.
Because of these factors the Schiff moment of $^{221}$Fr in the
{\em laboratory} frame is about 12 times less than that of  $^{223}$Fr,
although in the {\em intrinsic} frame the values of $S_{\text{intr}}$ 
are roughly equal.

There is a controversy regarding the g.s.\ spin and parity doublet in
$^{229}$Pa, which is on the border of the region of octupole deformed
nuclei. Two assignments: {\em a)}~\mbox{$K=\frac{5}{2}$},
\mbox{$I=\frac{5}{2}$} g.s.\ and 220~eV energy splitting within the
parity doublet \cite{Ahmad-Butler,Sheline-229Pa} 
and {\em b)}~\mbox{$K=\frac{1}{2}$}, \mbox{$I=\frac{3}{2}^{-}$}
g.s.\ and unidentified parity partner level 
\cite{Levon-229Pa,Butler-Nazarewicz-RMP} were made.
In case {\em a)} our calculations give using the experimental value
of the energy splitting the admixture coefficient
\mbox{$\alpha=640\times 10^{7}\eta$} 
and the Schiff moment in the laboratory frame,  
\mbox{$S=230000\times 10^{-8}\eta~e~\text{cm}$}.
Note however that in Table \ref{moments-table-big} the results for 
$^{229}$Pa are an order of magnitude smaller because a theoretical
value of 5~keV was used for the energy splitting of the doublet.

\subsection{Calculation of Atomic Electric Dipole Moments}
\label{subsec:calc-atomic}
 
Atomic electric dipole moment can be calculated using 
Eq.\ (\ref{eaedm3a}).  However, we do not need new complicated 
numerical calculations to find the EDM of interest. We can use numerous
calculations for the lighter atomic analogs (Xe, Hg, Cs) and introduce
corresponding corrections taking into account the $Z$-dependence of
the effect to find EDM of heavy atoms (Rn, Ra and Fr correspondingly).
Indeed, it follows from the atomic calculations that atomic EDM in
Eq.\ (\ref{eaedm3a}) is saturated by the contributions of 
electrons from the external shells which are similar in the analogous
atoms (the energy denominators for the transitions from these shells
are small and radial integrals are large). 
The expression for the atomic EDM is a product of
three factors: matrix elements of the radius 
\mbox{$\langle k_1 | r_z | k_2\rangle$},
energy denominators \mbox{$E_{k_1} - E_{k_2}$} and matrix elements
of the \mbox{T-}, P- odd nuclear electric potential 
\mbox{$\langle k_1 | -e \varphi^{(3)} | k_2\rangle$}.
The first two factors are determined by the wave functions at large
distances and they are the same in analogous atoms.  This fact is
deduced from numerous semi-empirical and computer calculations, from
experimental data for energy levels and probabilities of
electromagnetic transitions, as well as from the data on atomic
polarizabilities for analogous atoms (the expression for the
polarizability also contains radial integrals and energy
denominators).  The matrix elements
\mbox{$\langle k_1 | -e \varphi^{(3)} | k_2 \rangle$}
are determined by the wave function at small distances
(more accurately, by the gradient of the external electron density at
the nucleus) which strongly depends on the nuclear charge. However,
this dependence was calculated analytically
\cite{Sushkov-Flamb-Khripl}. The contribution of the Schiff moment is
determined by the matrix element between $s_{1/2}$ and $p_{1/2}$
(or $s_{1/2}$ and $p_{3/2}$)
electron orbitals which is proportional to \mbox{$SZ^{2}R_{1/2}$}
(or \mbox{$SZ^{2}R_{3/2}$}). The number of $p_{3/2}$ states is two
times larger then the number of $p_{1/2}$, therefore we will need
linear combination of the relativistic factors 
$R_{sp}=(R_{1/2}+2R_{3/2})/3$.

The relativistic factors $R_{1/2}$ and $R_{3/2}$  are given by
\begin{eqnarray}
&&R_{1/2}=\frac{4\gamma_{1/2}}{[\Gamma(2\gamma_{1/2}+1)]^{2}}
\biggl (\frac{2ZR_{0}}{a_{_{B}}} \biggr )^{2\gamma_{1/2}-2} \ ,
\nonumber \\
&&R_{3/2}=\frac{48}{\Gamma(2\gamma_{1/2}+1)\Gamma(2\gamma_{3/2}+1)}
\biggl (\frac{2ZR_{0}}{a_{_{B}}} 
\biggr )^{\gamma_{1/2}+\gamma_{3/2}-3} \ ,  \nonumber \\
\label{fb-at13}
\end{eqnarray}
where
\mbox{$\gamma_{j}=[(j+\mbox{$\frac{1}{2}$})^{2}-(Z\alpha)^{2}]^{1/2}$},
$a_{_{B}}$ is the Bohr radius and $R_{0}$ is the nuclear radius. In
light atoms $R_{sp} \simeq 1$, in heavy atoms $R_{sp} \simeq 10$.
Thus, we have simple estimates of the electric dipole moments of Ra,
Rn and Fr induced by the nuclear Schiff moments:
\begin{eqnarray}
&&d_{\text{at}}({\text{Ra}})=d_{\text{at}}(\text{Hg})
\frac{(SZ^{2}R_{sp})_{\text{Ra}}}{(SZ^{2}R_{sp})_{\text{Hg}}} \ ,
\nonumber \\
&&d_{\text{at}}({\text{Rn}})=d_{\text{at}}(\text{Xe})
\frac{(SZ^{2}R_{sp})_{\text{Rn}}}{(SZ^{2}R_{sp})_{\text{Xe}}} \ ,
\nonumber \\
&&d_{\text{at}}({\text{Fr}})=d_{\text{at}}(\text{Cs})
\frac{(SZ^{2}R_{sp})_{\text{Fr}}}{(SZ^{2}R_{sp})_{\text{Cs}}} \ ,
\label{fb-A1}
\end{eqnarray}
where only the nuclear Schiff moment contributions to atomic EDMs
$d_{\text{at}}$ of Hg, Xe and Cs are taken into account.
The atomic structure ratios here are
\begin{eqnarray}
&&
\frac{(Z^{2}R_{sp})_{\text{Ra}}}{(Z^{2}R_{sp})_{\text{Hg}}}=1.6 \ ,
\nonumber \\
&&
\frac{(Z^{2}R_{sp})_{\text{Fr}}}{(Z^{2}R_{sp})_{\text{Cs}}}\simeq
\frac{(Z^{2}R_{sp})_{\text{Rn}}}{(Z^{2}R_{sp})_{\text{Xe}}}=7.7 \ .
\label{relrat} 
\end{eqnarray}
The calculation of EDM of Hg, Xe and Cs have been done in
Refs.\ \cite{Sushkov-Flamb-Khripl,Dzuba-Flambaum,FKS-86}. 
The results of our calculations 
for Ra, Rn and Fr are presented in the Table \ref{moments-table-big}.

\section{SUMMARY}
\label{sec:summary-pt}

In this paper we studied of the T-odd, P- odd electric moments in
heavy nuclei with intrinsic reflection asymmetry and induced electric
dipole moments in corresponding atoms.  We presented a detailed theory
of the collective \mbox{T-,} P- odd electric moments in reflection
asymmetric odd-mass nuclei, in particular the Schiff moment.  
We employed the two-fluid liquid-drop model, particle plus core model,
and used the results of Nilsson-Strutinsky mean field calculations for
intrinsic reflection asymmetric nuclear shapes.  Various corrections
for nuclear T-odd, P-odd electric moments were evaluated.

In the calculations of induced atomic electric dipole moments
we employed the scaling relations between such moments in heavy atoms
and their lighter analogs and used the results of the calculations
for the latter
to find the corresponding moments in heavy atoms.
We studied the cases of all the heavy reflection-asymmetric odd-mass
nuclei for which there is evidence of intrinsic octupole
deformation in the ground state and which are relatively long-lived,
so their atoms could be suitable for experiment.

The results can be summarized in the following form:
\\
{\em a)}  In a reflection asymmetric nucleus which 
has odd mass number or is odd-odd, enhanced
collective T-odd, P-odd electric moments appear, if T-odd, P-odd
terms are present in the nuclear Hamiltonian.
\\
{\em b)}  The T-odd, P-odd Schiff moments in heavy nuclei with
intrinsic reflection asymmetry are typically
enhanced by more than two orders of magnitude in comparison with
reflection-symmetric deformed or spherical nuclei. 
\\
{\em c)}  Due to the atomic structure effects, atomic
electric dipole moments in heavy atoms are enhanced, compared to the
lighter analogs.
For atoms of nuclei with $Z$ around 90, the
atomic enhancement is of about 8 times,
in comparison with analog atoms with $Z$ around 55.
This enhancement factor is about 2 compared to analogs with
$Z$ around 80.
\\
{\em d)}  The atomic electric dipole moments, induced by T-odd, P-odd
hadron-hadron interaction in the nuclei studied are typically
enhanced 400--1000 times in comparison with Hg and Xe nuclei, for which
the best experimental upper limits on atomic electric dipole moments
are obtained. These findings may open new experimental possibilities
of studying time reversal violation.

\acknowledgments
We wish to thank J.D.~Bowman and I.B.~Khriplovich for discussions.
This work was supported by the US-Israel
Binational Science Foundation and by a grant for Basic Research of the
Israel Academy of Science. 

\appendix
\label{app:schiff-theorem}

\section{Screened \mbox{T-,} P- odd electrostatic potential of a
nucleus and the Schiff theorem}

The Hamiltonian of an atom placed in a homogeneous external electric
field $\bbox{E}_{0}$ is
\begin{eqnarray}
H&=&\sum_{i} 
\Bigl(K_{i}-e\varphi_{0}(\bbox{R}_{i})-e\bbox{R}_{i}\bbox{E}_{0}
\Bigr)
\nonumber \\
&+& \sum_{i>k} \frac{e^{2}}{\vert \bbox{R}_{i}-\bbox{R}_{k}\vert}-
\bbox{d}\bbox{E}_{0} \ ,
\nonumber \\
\varphi_{0}(&\bbox{R}_{i}&)=e\int 
\frac{\rho(\bbox{r})d^{3}\bbox{r}}{\vert\bbox{R}_{i}-\bbox{r}\vert} \ .
\label{astm-1}
\end{eqnarray}
Here $K_{i}$ and $\bbox{R}_{i}$ are the kinetic energy and coordinate
of the electron, $\varphi_{0}(\bbox{R}_{i})$
is the electrostatic nuclear potential and $\bbox{d}$ is the nuclear
dipole moment. 
Let us add to $H$ an auxiliary term
\begin{equation}
V=\bbox{d}\bbox{E}_{0}-\frac{1}{eZ}
\sum_{i}\bbox{d}\bbox{\nabla}_{i}\varphi_{0}(\bbox{R}_{i}) \ .
\label{astm-2}
\end{equation}
It is easy to demonstrate that in the linear approximation in
$\bbox{d}$ the interaction $V$ does not produce any energy shift,
\mbox{$\langle V\rangle =0$}. Indeed
\begin{equation}
\frac{i}{m}\Bigl [ \sum_{i}\bbox{p}_{i},H\Bigr ]=
-e\sum_{i}\bbox{\nabla}_{i}\varphi_{0}(\bbox{R}_{i}) 
+ Ze\bbox{E}_{0}  \ .
\label{astm-3}
\end{equation}
We have taken into account that the total electron momentum
\mbox{$\sum_{i}\bbox{p}_{i}$} commutes with the electron-electron
interaction term. Using Eq.\ (\ref{astm-2}) and 
$\mbox{$\langle n \vert [H,\sum_{i}\bbox{p}_{i}]\vert n\rangle$}
\sim \mbox{$(E_{n}-E_{n})$}=0$  we obtain
\mbox{$\langle V\rangle=
\bbox{d}\bbox{E}_{0}-\frac{1}{eZ}eZ\bbox{d}\bbox{E}_{0}=0$}.

To find an electric dipole moment one needs to measure a linear energy
shift in an external electric field. Since $V$ does not contribute to
this shift we can add it to the Hamiltonian
\begin{eqnarray}
\tilde{H}\equiv H+V&=&\sum_{i} 
\Bigl(K_{i}-e\varphi(\bbox{R}_{i})-
e\bbox{R}_{i}\bbox{E}_{0}\Bigr ) \nonumber \\
&+& \sum_{i>k} \frac{e^{2}}{\vert \bbox{R}_{i}-\bbox{R}_{k}\vert} \ ,
\nonumber \\
\varphi(\bbox{R}_{i})=\varphi_{0}(\bbox{R}_{i})&+&
\frac{1}{eZ}\bbox{d}\bbox{\nabla}_{i}\varphi_{0}(\bbox{R}_{i}) \ .
\label{astm-4}
\end{eqnarray}
Note, that the Hamiltonian $\tilde{H}$ does not contain the direct
interaction $\bbox{d}\bbox{E}_{0}$ between the nuclear electric dipole
moment and external field (Schiff theorem
\cite{Purcell-Ramsey,Schiff}). The dipole term is also canceled out in
the multipole expansion of $\varphi(\bbox{R}_{i})$.

%%%%%%%%

\begin{table}
\vspace*{-0.2in}
\caption{Intrinsic g.s.\ deformations
and energy splittings between opposite parity core states.}
\begin{tabular}{lccccccc}
           &$^{223}$Ra&$^{225}$Ra&$^{223}$Rn&
$^{221}$Fr & $^{223}$Fr &$^{225}$Ac &$^{229}$Pa \\
\hline
$\beta_{2}$& 0.125 & 0.143 & 0.129 & 0.106 & 0.122 & 0.138 &0.176 \\
$\beta_{3}$& 0.100 & 0.099 & 0.081 & 0.100 & 0.090 & 0.104 &0.082 \\
$\beta_{4}$& 0.076 & 0.082 & 0.078 & 0.069 & 0.076 & 0.078 &0.093 \\
$\beta_{5}$& 0.042 & 0.035 & 0.024 & 0.045 & 0.033 & 0.038 &0.020 \\
$\beta_{6}$& 0.018 & 0.016 & 0.023 & 0.020 & 0.022 & 0.013 &0.015 \\
$E_{c}$ (keV) & 212 & 221  & 213   &  305  &  212  &  206  & 333 \\
\end{tabular}
\label{deformations-table}
\end{table}

%%%%

\mediumtext
\begin{table}
\caption{Admixture coefficients $\alpha$ (absolute values)
and theoretical energy splitting between the g.s.\ doublet levels
\mbox{$\Delta E=E^{-}-E^{+}$},
parities of the intrinsic (reflection-asymmetric) 
single-particle g.s.\, 
calculated using the Woods-Saxon (W) and Nilsson (Nl) potentials,
experimental energy splitting,
{\em intrinsic} Schiff moments and Schiff moments 
(with Woods-Saxon potential)
as well as induced atomic dipole moments.
The values for $^{199}$Hg, $^{129}$Xe and $^{133}$Cs from 
 Refs.\ \protect\cite{FKS-86,Dzuba-Flambaum,Sushkov-Flamb-Khripl} 
are given for comparison.}
\begin{tabular}{lcccccccccc}
                     &$^{223}$Ra&$^{225}$Ra&$^{223}$Rn&
$^{221}$Fr&$^{223}$Fr&$^{225}$Ac& 
$^{229}$Pa&$^{199}$Hg&$^{129}$Xe&$^{133}$Cs \\
\hline
$\alpha\text{(WS)}~ 
(10^{7}~\eta)$&1.   & 2.  & 4. &
0.7   &2. & 3.  & 34.  \\
$\Delta E\text{(WS)}$ (keV) 
                 & 170. & 47.  & 37. &
216. &75.  & 49. & 5. \\
$\pi_{\text{p}}$(WS) 
           & 0.81 & -0.02 & 0.17 & -0.55 & -0.34 &  -0.35& 0.01 \\ 
\hline
$\alpha\text{(Nl)}~
(10^{7}~\eta)$&2. & 5.  & 2.  & &
 &   &    \\
$\Delta E\text{(Nl)}$~(keV) 
                  &171. &  55. & 137. &  &
   &   &  & \\
\hline
$\Delta E_{\text{exp}}$ (keV) 
& 50.2& 55.2 & & 234.& 160.5& 40.1& 0.22& &\\
$S_{\text{intr}}~
(e~\text{fm}^{3})$&  24   & 24    &   15 & 21 & 20 & 28 &25   &\\
$ $S$~ 
\mbox{$(10^{8}~\eta~e~\text{fm}^{3})$}$
                & 400& 300& 1000& 43 & 500 & 900 &
1.2$\times 10^{4}$&   -1.4 &  1.75 &  3 \\    
$\mbox{$d(\text{at})$}~ 
\mbox{$(10^{25}~\eta~ e~\text{cm})$}$
                & 2700& 2100& 2000 & 240 & 2800 &  &
 &      5.6&   0.47 & 2.2   \\ 
\end{tabular}
\label{moments-table-big}
\end{table}
\narrowtext   

\end{document}